\documentclass[aps,prb,reprint,showpacs,superscriptaddress]{revtex4-1}
\usepackage{amsmath,times}
\usepackage{graphicx}
\usepackage{bm}
\begin{document}
\title{Quantum computing through electron propagation in the edge states of quantum spin Hall systems}
\author{Wei Chen}
\affiliation{National Laboratory of Solid State Microstructures and Department of Physics, Nanjing University, Nanjing 210093, China}
\affiliation{Department of Physics and Center of Theoretical and Computational Physics, The University of Hong Kong, Pokfulam Road, Hong Kong, China}
\author{Zheng-Yuan Xue}
\affiliation{Department of Physics and Center of Theoretical and Computational Physics, The University of Hong Kong, Pokfulam Road, Hong Kong, China}
\affiliation{Laboratory of Quantum Information Technology, and School of Physics and Telecommunication Engineering, South China Normal University, Guangzhou 510006, China}
\author{Z. D. Wang}
\email{zwang@hku.hk}
\affiliation{Department of Physics and Center of Theoretical and Computational Physics, The University of Hong Kong, Pokfulam Road, Hong Kong, China}
\author{R. Shen}
\email{shen@nju.edu.cn}
\affiliation{National Laboratory of Solid State Microstructures and Department of Physics, Nanjing University, Nanjing 210093, China}
\author{D. Y. Xing}
\affiliation{National Laboratory of Solid State Microstructures and Department of Physics, Nanjing University, Nanjing 210093, China}
\begin{abstract}
We propose to implement quantum computing based on electronic spin qubits by controlling the propagation of the electron wave packets through the helical edge states of quantum spin Hall systems (QSHs). Specifically, two non-commutative single-qubit gates, which rotate a qubit around $z$ and $y$ axes, can be realized by utilizing gate voltages either on a single QSH edge channel or on a quantum point contact structure. The more challenging two-qubit controlled phase gate can be implemented through the on-demand capacitive Coulomb interaction between two adjacent edge channels from two parallel QSHs. As a result, a universal set of quantum gates can be achieved in an all-electrical way. The fidelity and purity of the two-qubit gate are calculated with both time delay and finite width of the wave packets taken into consideration, which can reach high values with the existing high-quality single electron source.
\end{abstract}
\maketitle

\section{introduction}
Topological insulators are new quantum states of matter discovered in recent years, \cite{Kane,Zhang} which are characterized by fully gapped bulk states and gapless edge or surface states protected by the band topology. These edge or surface states can be well described by the Dirac equations, which leads to several remarkable results. First, the spin-momentum locking suggests their potential applications in spintronics \cite{Qi,Richter} and quantum information processes based on electronic spins. \cite{Chen,Chen2} Second, when a superconducting gap is induced by proximity to a conventional s-wave superconductor, the superconducting phases are predicted to be topologically nontrivial as well, which may host non-Abelian Majorana fermions. \cite{Fu,Fu2} Topological quantum computation based on the Majorana fermions can achieve fault-tolerance at the physical level, which has become one of the most exciting approaches to realize a full-scale quantum computer. \cite{Kitaev, Sarma}

In this paper, we propose to perform quantum computing based on electronic spin states \cite{Loss} in the two-dimensional topological insulators, also known as quantum spin Hall systems (QSHs), which has been realized in both HgTe/CdTe \cite{Molenkamp} and InAs/GaSb \cite{Knez,Knez2} quantum wells. The computing processes are achieved by on-demand controlling the transportation of wave packets (WPs) through the helical edge channels of QSHs. It is shown that a universal set of quantum gates can be achieved simply by the gate voltage control. Specifically, a phase gate is implemented by depositing a side gate on either edge channel of the QSH, with the phase shift being controlled by the gate voltage. A single-qubit rotation around the $y$ axis can be realized by adjusting the tunneling of electron between edges in a quantum point contact structure. A two-qubit controlled phase gate can be constructed by two parallel QSHs, where two adjacent edge channels from different QSHs are coupled via on-demand capacitive Coulomb interaction. With the help of recent developed high-quality single electron source, the fidelity of the gate can reach a high value. Moreover, given that braiding operations on the Majorana fermions are still not sufficient for the implementation of universal quantum computation,\cite{Sarma} our proposal may be utilized as the ancillary qubits, \cite{Beenakker,Sau,Jiang,Bonderson,Flensberg} in order to realize hybrid quantum computation with a high error threshould \cite{Bravyi} solely within the framework of QSHs.

The rest of this paper is organized as follows. The implementation of the single-qubit gates are introduced in Sec. \ref{II}, and the two-qubit controlled phase gate is studied in Sec. \ref{III}. The effect on the fidelity and purity of the two-qubit gate due to the delay between electron pulses and the spread of the WPs is studied in Sec. \ref{IV}. Finally, a brief discussion is presented in Sec. \ref{VI}.

\section{single-qubit gates}\label{II}
\begin{figure}
\centering
\includegraphics[width=0.48\textwidth]{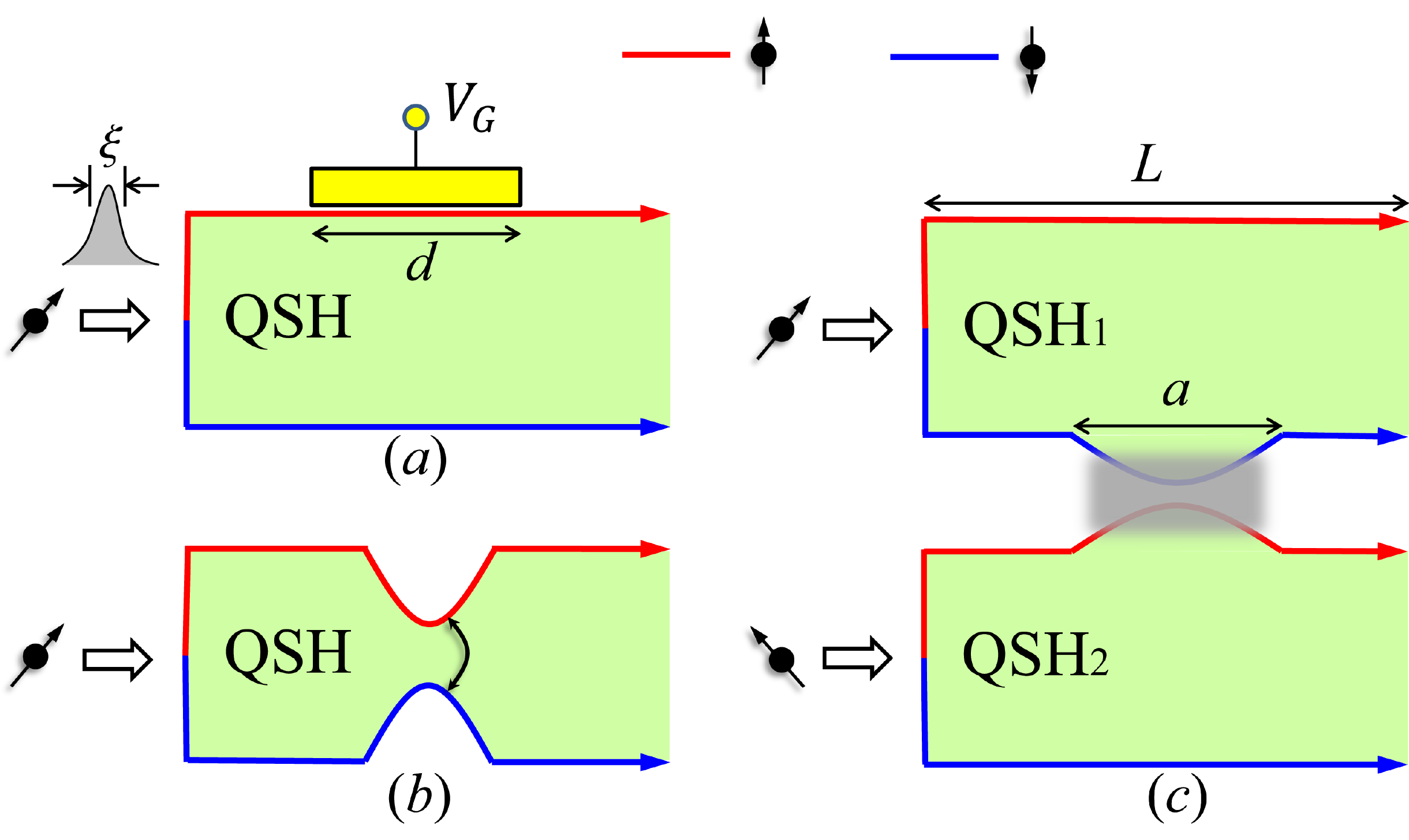}
\caption{(color online) Scheme of quantum computing in quantum spin Hall systems (QSHs), where two edge channels corresponding to opposite spins are sketched. The Lorentzian type electron wave packets with a typical size $\xi$ are injected into the edge channels of the QSHs and undergo certain spin manipulations. (a) A phase shift gate is realized by a gate voltage $V_g$ applied on the upper edge; (b) A spin rotation around $y$ axis is achieved by tunneling between two edges in a quantum point contact structure; (c) A two-qubit controlled phase gate is achieve via capacitive Coulomb interaction (shadow area) between two adjacent edge channels from two parallel QSHs.} \label{fig1}
\end{figure}

Our implementation of a universal set of the quantum gates is sketched in Fig. \ref{fig1}. Two single-qubit gates are both composed by one piece of QSH as shown in Fig. \ref{fig1}(a,b), while the two-qubit gate is constructed by two parallel QSHs as shown in Fig. \ref{fig1}(c). We assume electron pulses are injected into the edge channels at $x=-L/2$, and transport along the one dimensional edge channels to undergo certain computing operations, until they finally reach $x=L/2$.

For the implementation of quantum computing, on-demand single electron sources are required, which can be generated with the help of suitable voltage pulses. \cite{Ivanov, Feve, Feve2} In the following, we consider an incoming WP with the Lorentzian shape \cite{Ivanov} initially located at $x=-L/2$, which has the form of $f(x)=\sqrt{\frac{\xi}{\pi}}\frac{1}{x+L/2+i\xi}$ with $\xi$ being the the width of the WP. The entire wave function including the spin part can be expressed as
\begin{equation}\label{s0}
|\Psi(0)\rangle=\int dxf(x)\sum_\sigma \alpha_\sigma\psi_\sigma^\dag(x)|0\rangle,
\end{equation}
where $(\alpha_\uparrow, \alpha_\downarrow)^T$ is the initial spin state, $\psi_\sigma^\dag(x)$ creates an electron at $x$ with its spin $\sigma$ taking the values of $\uparrow$ or $\downarrow$, and $|0\rangle$ represents the vacuum state at zero temperature.

The helical electrons in the edge states can be well described by the massless Dirac equation, and the back scattering within one edge is completely suppressed even when nonmagnetic impurities and potential barriers exist. Here, we focus on the right moving electrons only, and the edge channels are sketched in Fig. \ref{fig1}, which can be described by the Hamiltonian
\begin{equation}\label{h0}
H_0=-i\hbar v\sum_\sigma\int\psi_\sigma^\dag(x)\partial_x\psi_\sigma(x)dx,
\end{equation}
where $v$ is the Dirac velocity. Due to the spin-momentum locking, electrons with spin $\uparrow$ and $\downarrow$ transport rightward in the upper and lower edge channels, respectively. The spatial separation of opposite spins allows one to manipulate the spin states by applying electrical potential on the edge channels, which is easier to implement compared with the magnetic method.

\subsection{phase gate}

The phase gate
$U_z(\varphi)=\mathrm{diag}\{e^{-i\varphi}, 1\}$
operating on the spin state $(\alpha_\uparrow, \alpha_\downarrow)^T$ can be achieved by depositing a side gate on the upper edge channel as shown in Fig. \ref{fig1}(a), which can be expressed by the gating Hamiltonian
\begin{equation}\label{hg}
H_g=eV_g\int^{d/2}_{-d/2}\psi^\dag_\uparrow(x)\psi_\uparrow(x)dx,
\end{equation}
with $V_g$ being the gate voltage, applied on the region $(-d/2,d/2)$.

Such a side gate configuration has been studied in the interferometer structures in QSHs by the plane-wave Ansatez. \cite{Dolcini, Sassetti} Here, by contrast, we focus on the propagation of a spatially localized WP in Eq. (\ref{s0}), for the purpose of on-demand control over the electronic spin, which is not an eigenstate of the total Hamiltonian of Eq. (\ref{h0}, \ref{hg}). Interestingly, it turns out that, localized WP of massless Dirac electron can transport without any dispersion, just as the light travels in vacuum. One can see this directly through the equation of motion of the field operator in the interaction picture $\psi_\sigma(x,t)=e^{-\frac{H_0t}{i\hbar}}\psi_\sigma(x)e^{\frac{H_0t}{i\hbar}}$,
\begin{equation}
i\hbar\frac{\partial}{\partial t}\psi_\sigma(x,t)=[\psi_\sigma(x,t), H_0].
\end{equation}
Solving the equation then leads to the result
\begin{equation}
\psi_\sigma(x,t)=e^{-vt\partial_x}\psi_\sigma(x)=\psi_\sigma(x-vt),
\end{equation}
which says that for any kind of WP, it can transport along the edge channels with a velocity $v$, while keeping its shape unchanged.

We consider that at $t=0$, an incoming WP in the form of Eq. (\ref{s0}) is located around $x=-L/2$. In the interaction picture, the gating Hamiltonian is $H_g^i(t)=eV_g\int^{d/2}_{-d/2}\psi_\uparrow^\dag(x-vt)\psi_\uparrow(x-vt)dx$, which is now dependent on time. The wave function in the interaction picture can be calculated as well under the evolution of the gating Hamiltonian by $|\Psi^i(t)\rangle=e^{\frac{1}{i\hbar}\int_0^tH_g^i(t')dt'}|\Psi(0)\rangle$, and the result is
\begin{equation}
\begin{split}
|\Psi^i(t)\rangle&=\int dxf(x)\sum_\sigma\alpha_\sigma e^{-i\varphi_\sigma^i(t)}\psi^\dag_\sigma(x)|0\rangle,\\
\varphi^i_\sigma(t)&=\delta_{\sigma,\uparrow}\frac{eV_g}{\hbar}\int_0^tdt'\int_{-d/2}^{d/2}dx'\delta(x'-vt'-x).
\end{split}
\end{equation}
Finally, the wave function in the Schr\"{o}dinger picture can be obtained through the relation $|\Psi(t)\rangle=e^{\frac{H_0t}{i\hbar}}|\Psi^i(t)\rangle$, or specifically,
\begin{equation}\label{s1}
\begin{split}
|\Psi(t)\rangle&=\int dxf(x-vt)\sum_\sigma\alpha_\sigma e^{-i\varphi_\sigma(t)}\psi^\dag_\sigma(x)|0\rangle,\\
\varphi_\sigma(t)&=\delta_{\sigma,\uparrow}\frac{eV_g}{\hbar}\int_0^tdt'\Theta(vt'-L/2+d/2)\Theta(L/2+d/2-vt'),
\end{split}
\end{equation}
where $\Theta$ is the step function, and the width of the WP is assumed to be narrow compared with the scale of the QSH, i.e., $\xi\ll L$. When the electron arrives at $x=L/2$ after a traveling time of $\tau_0=L/v$, the additional phase accumulated in the upper channel is $\varphi(\tau_0)=eV_gd/\hbar v$, provided that $L-d\gg \xi$. As a result, the spin state can be expressed as $(e^{-i\varphi}\alpha_\uparrow, \alpha_\downarrow)^T$ after tracing out the spatial part of the wave function in Eq. (\ref{s1}), which manifests that the side gate can be utilized as a phase gate on electronic spin qubit.

More practically, taking the InAs/GaSb quantum wells as an example, \cite{Knez} where the Dirac velocity is $v=3\times10^4$m/s, so a $\pi/8$-phase gate can be achieved by setting the length of the gating region to $d=7.8$nm and applying a gate voltage of $V_g=1$mV. Given that the phase coherence length of the edge channels in the InAs/GaSb quantum wells is $l_\phi\simeq2\mu \mathrm{m}\gg d$, it is reasonable to expect a high quality of the gate, for the decoherence effect is negligible.

\subsection{rotation gate}

To perform universal quantum computing, we now further consider an operation which rotates the spin around the $y$ axis. Such a quantum gate can be realized in a quantum point contact structure \cite{Teo} as shown in Fig. \ref{fig1}(b), and it has been utilized in several quantum information processing tasks based on electronic spins. \cite{Chen,Chen2} According to the study done by Krueckl and Richter, \cite{Richter} there exists an energy window in HgTe/CdTe quantum wells, where the amplitudes of electron scattering between edges can be adjusted to full range by a top gate $V_g^T$, with the back scattering being completely suppressed simultaneously. This indicates that, within such an energy widow, the quantum point contact structure can serve as an ideal spin dependent beam splitter, which is just suitable for quantum computing. The rotation operation on spin around the $y$ axis by an angle $\theta$ can be expressed effectively by $U_y(\theta)=\cos(\theta/2)-i\sin(\theta/2)\sigma_y$, where $\theta$ is a function of the gate voltage and $\sigma_y$ is the Pauli matrix. A detail correspondence between $\theta$ and the voltage applied on the top gate can be found in Ref. \onlinecite{Richter}. For example, when the Fermi energy takes the value $E_F=0$, then the rotation by an angle of $\theta=\pi$ can be achieved by tuning the gate voltage to the value $V_g^T=-4$meV. Though the calculations are performed in the HgTe/CdTe quantum wells,\cite{Richter} it is reasonable to expect a qualitative same conclusion in InAs/GaSb quantum wells,\cite{Knez,Knez2} given that the physics inside is independent on specific systems.

\section{controlled phase gate}\label{III}

The implementation of the single-qubit gates via electrical control as discussed in Sec. \ref{II} also suggest that a two-qubit gate can be achieved by Coulomb interaction between two electrons, which carry two spin qubits traveling through the edge channels. The structure is shown in Fig. \ref{fig1}(c), where a pair of QSHs parallel to each other are utilized. The adjacent edge channels of the two QSHs are coupled through the capacitive Coulomb interaction (the shadow area) in the interaction region $(-a/2, a/2)$, where the tunneling between edges is forbidden. This can be realized by the deformation of the QSH edges, or through the gate voltage control on the expansion of the edge states in the direction perpendicular the edge. It is intuitive that, when the upper electron picks the spin down channel and the lower one chooses the spin up channel at the same time, a joint phase due to the interaction energy is accumulated, which is absent for all the other three spin configurations. This suggests a two-qubit controlled phase gate can be implemented through the on-demand Coulomb interaction between electrons, which together with the two single-qubit gates can realize universal quantum computation. \cite{Nielsen} Such Coulomb interaction between edges has recently been proposed to measure the which-path information \cite{Dressel} and generate orbital entanglement \cite{Vyshnevyy} in the coupled Mach-Zehnder interferometer setup composed by the edge states in the quantum Hall systems. Here we generalize the discussion to QSHs and utilize such an effect to perform quantum gate on the electronic spins. The incoming electron wave now contains one electron in each QSH, and takes the form
\begin{equation}
\begin{split}
|\Phi(0)\rangle=&\int\int dx_1dx_2 f_1(x_1)f_2(x_2)\times\\
&\sum_{\sigma_1,\sigma_2}\beta_{\sigma_1\sigma_2}\psi_{1\sigma_1}^\dag(x_1)\psi_{2\sigma_2}^\dag(x_2)|0\rangle,
\end{split}
\end{equation}
where the subscripts 1, 2 label the upper and lower QSHs, respectively (Fig. \ref{fig1}c) and $(\beta_{\uparrow\uparrow},\beta_{\uparrow\downarrow},\beta_{\downarrow\uparrow},\beta_{\downarrow\downarrow})^T$ is a general two-spin state.

The capacitive Coulomb potential with an effective interacting length $a$ takes the form
\begin{equation}
H_{int}=V\int\int dx_1dx_2\kappa_1(x_1)\kappa_2(x_2)\rho_{1\downarrow}(x_1)\rho_{2\uparrow}(x_2),
\end{equation}
where $\rho_{j\sigma_j}(x_j)=\psi_{j\sigma_j}^\dag(x_j)\psi_{j\sigma_j}(x_j)$ is the density operator, $V$ is the interacting constant, and $\kappa_1(x)=\kappa_2(x)=\exp(-2|x|/a)$ are the interaction kernels. \cite{Dressel, Vyshnevyy, Blatter}

We now calculate the evolution of the two particle wave function. The interaction Hamiltonian in the interaction picture can be obtained as $H_{int}^i(t)=V\int\int dx_1dx_2\kappa_1(x_1+vt)\kappa_2(x_2+vt)\rho_{1\downarrow}(x_1)\rho_{2\uparrow}(x_2)$, with which the wave function in the interaction picture can be solved directly as
\begin{equation}
\begin{split}
|\Phi^i(t)\rangle=&\int\int dx_1dx_2 f_1(x_1)f_2(x_2)\times\\
&\sum_{\sigma_1,\sigma_2}\beta_{\sigma_1\sigma_2}e^{-i\phi^i_{\sigma_1\sigma_2}(t)}\psi_{1\sigma_1}^\dag(x_1)\psi_{2\sigma_2}^\dag(x_2)|0\rangle,\\
\phi^i_{\sigma_1\sigma_2}(t)=&\delta_{\sigma_1,\downarrow}\delta_{\sigma_2,\uparrow}\frac{V}{\hbar}\int_0^tdt'\kappa_1(x_1+vt')\kappa_2(x_2+vt').
\end{split}
\end{equation}
The Schr\"{o}dinger wave function is obtained via $|\Phi(t)\rangle=e^{\frac{H_0t}{i\hbar}}|\Phi^i(t)\rangle$ with $H_0$ now describing two QSHs and the final result is
\begin{equation}
\begin{split}
|\Phi(t)\rangle=&\int\int dx_1dx_2 f_1(x_1-vt)f_2(x_2-vt)\times\\
&\sum_{\sigma_1,\sigma_2}\beta_{\sigma_1\sigma_2}e^{-i\phi_{\sigma_1\sigma_2}(t)}\psi_{1\sigma_1}^\dag(x_1)\psi_{2\sigma_2}^\dag(x_2)|0\rangle,\\
\phi_{\sigma_1\sigma_2}(t)=&\delta_{\sigma_1,\downarrow}\delta_{\sigma_2,\uparrow}\frac{V}{\hbar}\int_0^td\tau\kappa_1(x_1-v\tau)\kappa_2(x_2-v\tau).
\end{split}
\end{equation}
The additional phase shared by both electrons are the function of coordinates $x_{1,2}$. After the electron pair traveling through the interaction region, the phase term has the following form
\begin{equation}\label{phase}
\phi_{\downarrow\uparrow}(\tau_0)=\phi_0e^{-\frac{2|x_1-x_2|}{a}}(1+\frac{2|x_1-x_2|}{a}),
\end{equation}
where the desired constant phase is determined by $\phi_0=Va/2\hbar v$.

We would like to investigate the ideal case first, as the width of the WPs is much smaller than the interacting length, i.e., $\xi\ll a$. This means the uncertainty of the electron positions are negligible, so as the uncertainty of the joint phase. As a result, the general two-qubit state after traveling through the interaction region can be expressed as $(\beta_{\uparrow\uparrow},\beta_{\uparrow\downarrow},e^{-i\phi_0}\beta_{\downarrow\uparrow},\beta_{\downarrow\downarrow})^T$, by tracing out the spatial part of the wave function. Therefore, a controlled phase gate $U_c$ is achieved by the on-demand Coulomb interaction, which has the form of
\begin{equation}
U_c=\mathrm{diag}\{1,1,e^{-i\phi_0},1\}.
\end{equation}

\section{fidelity and purity}\label{IV}
Practically, the WPs always have finite width $\xi$, and the injection of the wave pulse may not be perfectly synchronized as well, with a time delay of $\tau_d$. As a result, the spin state is generally described by the reduced density matrix $\hat{\rho}_R=\mathrm{Tr}_{x_1,x_2}\{|\Phi(t)\rangle\langle\Phi(t)|\}$, or explicitly,
\begin{equation}
\hat{\rho}_R=S\left(
                     \begin{array}{cccc}
                       1 & 1 & \Lambda^* & 1 \\
                       1 & 1 & \Lambda^* & 1 \\
                       \Lambda & \Lambda & 1 & \Lambda \\
                       1 & 1 & \Lambda^* & 1 \\
                     \end{array}
                   \right)S^*,
\end{equation}
with the matrix $S=\mathrm{diag}\{\beta_{\uparrow\uparrow},\beta_{\uparrow\downarrow},\beta_{\downarrow\uparrow},\beta_{\downarrow\downarrow}\}$, and the phase term being defined by
\begin{equation}
\begin{split}
\Lambda=\frac{1}{\pi^2}\int\int d\eta_1d\eta_2\frac{\exp[-i\phi_0 e^{-\lambda} (1+\lambda)]} {(\eta_1^2+1)(\eta_2^2+1)},
\end{split}
\end{equation}
where $\lambda=\nu|\eta_1-\eta_2+\Delta_d/\nu|$, with the dimensionless quantities $\Delta_d=2v\tau_d/a$ and $\nu=2\xi/a$ describing the delay between the wave pulses and the width of the WP, in scale of the screening length. The expression of $\Lambda$ suggests that it is not a simple exponential function with modulus 1, due to the finite width of the WP. The reason is that, the phase shift $\phi_{\downarrow\uparrow}(\tau_0)$ in Eq. (\ref{phase}) is a function of coordinates of both electrons. As a result, the probabilistic feature of the coordinates leads to the uncertainty of the phase $\phi_{\downarrow\uparrow}(\tau_0)$, or equivalently, spin decoherence, which is reflected by the suppression of the interference term.

\begin{figure}
\centering
\includegraphics[width=0.4\textwidth]{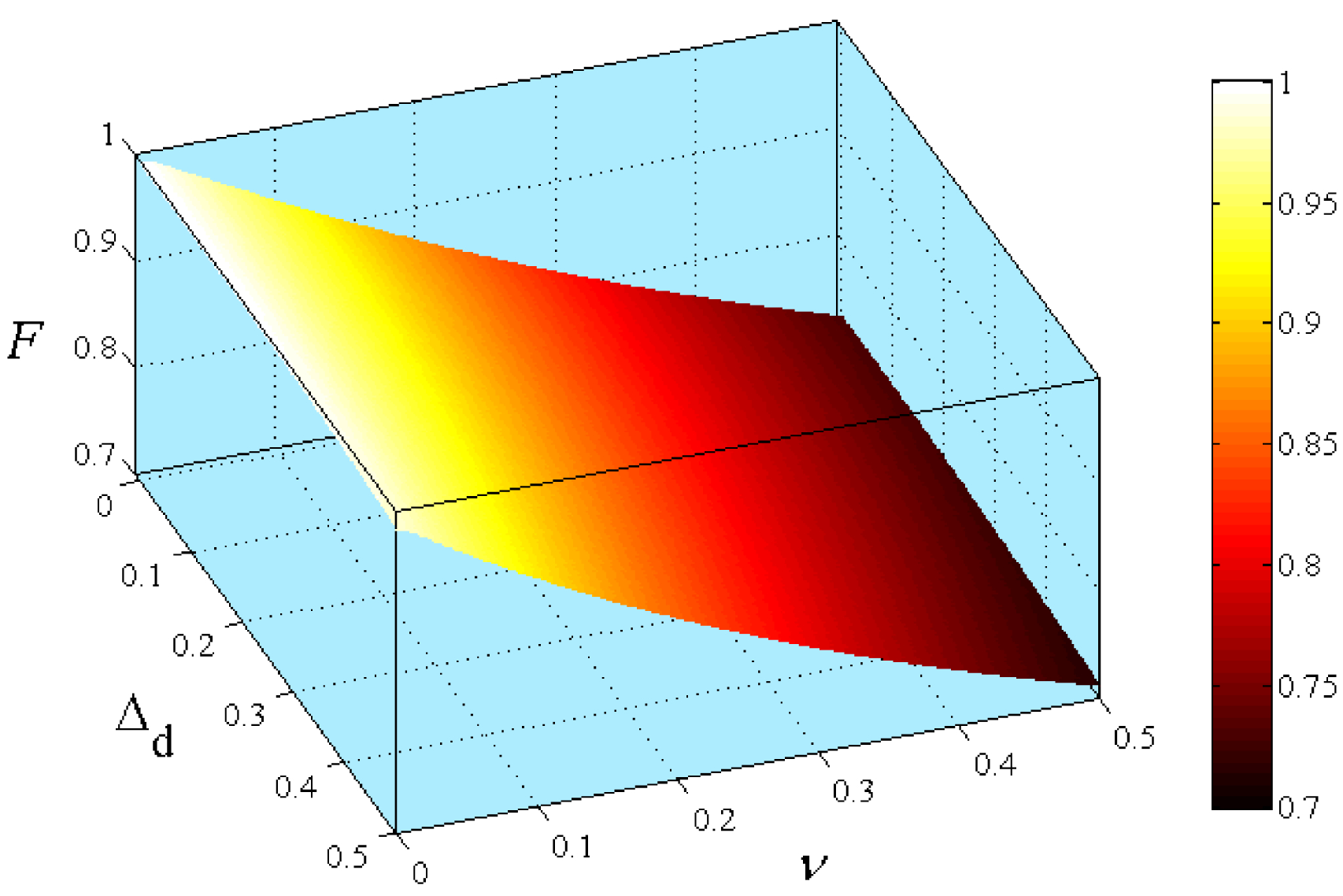}
\caption{(color online) The fidelity $\mathcal{F}$ as a function of $\Delta_d$ and $\nu$.} \label{fig2}
\end{figure}

\begin{figure}
\centering
\includegraphics[width=0.4\textwidth]{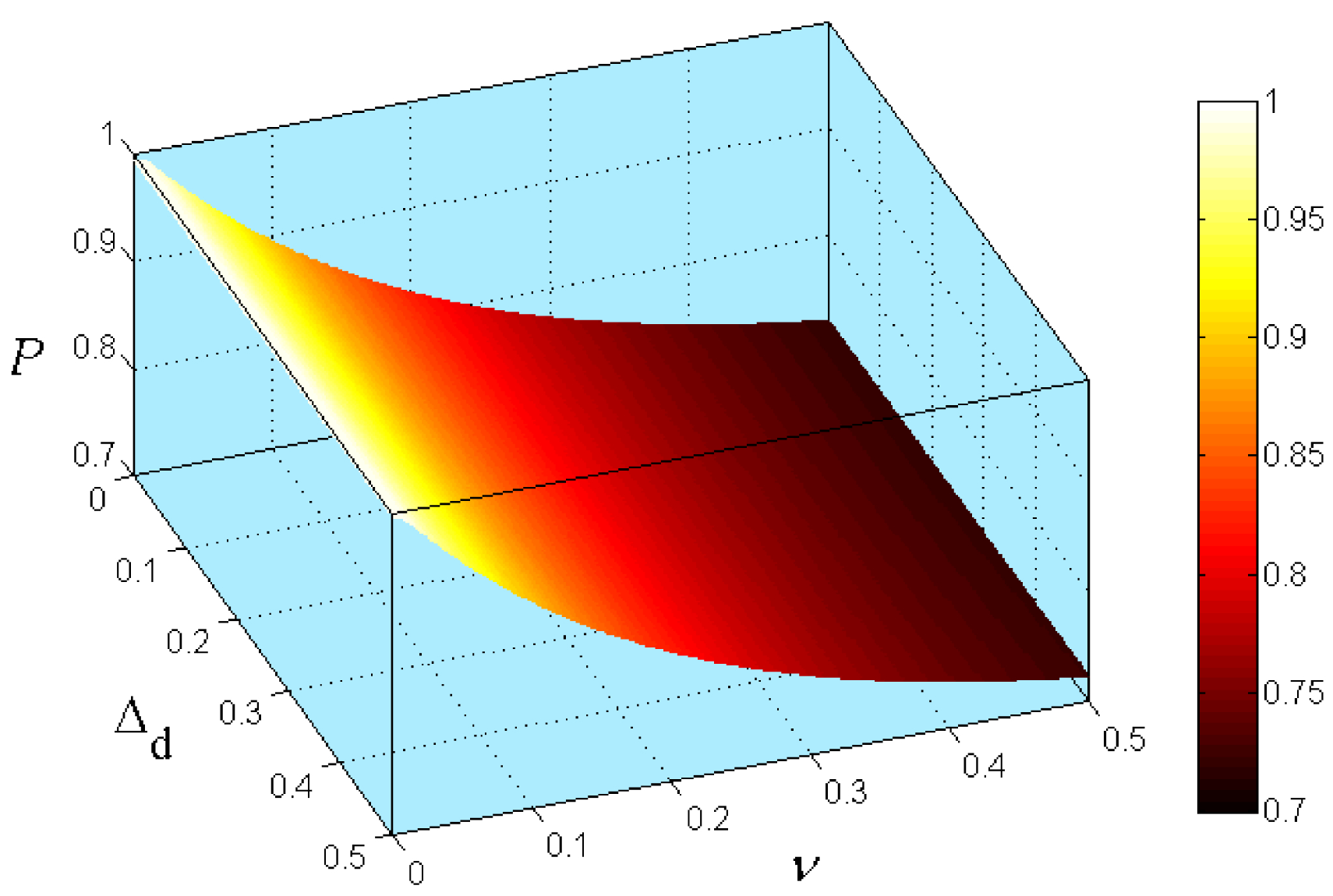}
\caption{(color online) The purity $\mathcal{P}$ as a function of $\Delta_d$ and $\nu$.} \label{fig3}
\end{figure}

To investigate the effect of the pulse delay and finite width of the WPs mentioned above, we assume $\phi_0=\pi$ in $U_c$ and the input two-spin state is $|\Phi_{in}\rangle=(1/2,1/2,1/2,1/2)^T$. The fidelity of the two-qubit gate can be defined by $\mathcal{F}=\langle\Phi_{in}|U_c^\dag\hat{\rho}_RU_c|\Phi_{in}\rangle$, which quantifies the closeness between the actually obtained final state and the target state, \cite{Zoller} and can be obtained as
\begin{equation}
\mathcal{F}=\frac{1}{8}(5-3|\Lambda|\cos\phi_c),
\end{equation}
where $\phi_c=\mathrm{arg}\{\Lambda\}$. The expression of $\mathcal{F}$ indicates that both the decoherence effect, which causes $|\Lambda|<1$, and an unexpected phase shift $\phi_c-\phi_0$, will reduce the fidelity.

Alternatively, we can also define the gate purity by $\mathcal{P}=\mathrm{Tr}\{(\hat{\rho}_R)^2\}$ to characterize the effects  of decoherence on the gate, \cite{Zoller} which has the form of
\begin{equation}
\mathcal{P}=\frac{1}{8}(5+3|\Lambda|^2),
\end{equation}
where we see that only the modulus of $\Lambda$ enters the expression.

Numerical results for both the fidelity $\mathcal{F}$ and purity $\mathcal{P}$ of the gate as a function of $\Delta_d$ and $\nu$ are shown in Fig. \ref{fig2} and Fig. \ref{fig3}, respectively. It turns out that the imperfectness of synchronization has a negligible effect compared with the finite width of the WPs, for the main effect of the former is the reduction of the presetting phase $\phi_0$. In contrast, the finite width of the WPs which introduces the decoherence effect has considerable impact on both the fidelity and purity. As a result, in order to realize a two-qubit gate with high fidelity, the width of the WPs is required to be sufficiently narrow.

Fortunately, recent progresses on the coherent single electron source shows the feasibility of our proposal. It has been reported that the emission time of a single electron pulse can reach the accuracy of picosecond, i.e., $\tau_\xi\sim1$ps, \cite{Feve2} which means the width of the WP in the edge channels of InAs/GaSb quantum wells \cite{Knez} may reach a value of $\xi=v\tau_\xi\sim30$nm. Therefore, by taking the interacting length of $a\sim600$nm, the fidelity and purity of the controlled phase gate can reach the values of 93.3\% and 89.0\%, respectively.

In the above calculation, the inherent decoherence effect \cite{Recher} due to the interaction with the Fermi see is assumed to be weak. \cite{Dolcini} Practically, the strength of the screened Coulomb interaction depends on the geometrical parameters of the quantum wells, which can range from the weakly interacting to strongly interacting limit. \cite{Teo,Zhang2} It is also worth noting that, the decoherence caused by Coulomb interaction in the Luttinger liquids can be partly or even fully undone with the help of a suitable voltage pulse. \cite{Vyshnevyy,Blatter} Consequently, the quantum computing scheme based on electronic spins seems rather promising within the QSH framework.

\section{discussion}\label{VI}

We have shown that a universal set of quantum gates on electronic spins can be realized by electrically controlled WP propagation in the edge channels of QSHs. It is promising to reach satisfactory fidelity and purity of the quantum gates based on the existing experimental techniques, such as the high-quality single electron sources. \cite{Feve2} We find that InAs/GaSb quantum wells is a better candidate for the implementation of quantum computing compared with HgTe/CdTe quantum wells, which has a slower Dirac velocity and a longer phase coherence length. \cite{Knez} A braiding proposal in QSHs has been put forward recently, \cite{Mi} and it has been proved that the transfer between spin states and topological qubits is advisable due to the spin polarization of Majorana fermions. \cite{Flensberg} Therefore, the spin qubits can serve as the ancillary qubits to perform several quantum operations which can not be achieved by braiding, such as the $\pi/8$ and controlled phase gates. These conventional quantum gates combined with braiding operations over Majorana fermions may achieve universal quantum computation with a high error threshold in QSHs.

\begin{acknowledgments}
This work was supported by the GRF (HKU7058/11P), CRF (HKU-8/11G) of the RGC of Hong Kong and the URC fund of HKU, by 973 Program (Grants No. 2011CB922100, No. 2011CBA00205,  No. 2009CB929504, and No. 2013CB921804), by NSFC (Grants No. 11074111, No. 11174125, No. 11023002 and No. 11004065), by PAPD of Jiangsu Higher Education Institutions, by NCET, by the PCSIRT, and by the Fundamental Research Funds for the Central Universities.
\end{acknowledgments}

\end{document}